\documentclass{article}
\usepackage{cite}
\usepackage{authblk}
\usepackage{url}
\usepackage{graphicx}
\usepackage[a4paper, left=2.5cm, right=2.5cm, top=3cm, bottom=3cm]{geometry}
\usepackage{dcolumn}
\usepackage{bm}

\usepackage[utf8]{inputenc}
\usepackage[T1]{fontenc}
\usepackage{mathptmx}
\usepackage{algorithm}
\usepackage{algpseudocode}
\usepackage{amsmath, amssymb}
\usepackage{etoolbox}
\usepackage{multirow}
\usepackage{booktabs}
\usepackage{soul}
\usepackage{color}

\begin{document}

\title{Triadic Closure-Heterogeneity-Harmony GCN for Link Prediction} 

\author[1,*,\dag]{Ke-ke Shang}
\author[1,\dag]{Junfan Yi}
\author[2]{Michael Small}
\author[3,**]{Yijie Zhou}

\affil[1]{Nanjing University, Nanjing, China}
\affil[2]{The University of Western Australia, Perth, Australia}
\affil[3]{Communication University of Zhejiang, Hangzhou, China}

\date{}

\maketitle

\noindent {\small 
\textsuperscript{*} Corresponding author: Ke-ke Shang (@example.com)\\
\textsuperscript{**} Co-corresponding author: Yijie Zhou (@example.com)\\
\textsuperscript{\dag} These authors contributed equally to this work.}

\begin{abstract}
Link prediction aims to estimate the likelihood of connections between pairs of nodes in complex networks, which is beneficial to many applications from friend recommendation to metabolic network reconstruction.
Traditional heuristic-based methodologies in the field of complex networks typically depend on predefined assumptions about node connectivity, limiting their generalizability across diverse networks. While recent graph neural network (GNN) approaches capture global structural features effectively, they often neglect node attributes and intrinsic structural relationships between node pairs. To address this, we propose TriHetGCN, an extension of traditional Graph Convolutional Networks (GCNs) that incorporates explicit topological indicators---triadic closure and degree heterogeneity. TriHetGCN consists of three modules: topology feature construction, graph structural representation, and connection probability prediction. The topology feature module constructs node features using shortest path distances to anchor nodes, enhancing global structure perception. The graph structural module integrates topological indicators into the GCN framework to model triadic closure and heterogeneity. The connection probability module uses deep learning to predict links. Evaluated on nine real-world datasets, from traditional networks without node attributes to large-scale networks with rich features, TriHetGCN achieves state-of-the-art performance, outperforming mainstream methods. This highlights its strong generalization across diverse network types, offering a promising framework that bridges statistical physics and graph deep learning.
\end{abstract}

\section{Introduction}
\label{sec:level1}
Link prediction, a crucial problem in graph-structured research, aims to estimate the likelihood of the existence of a link between two nodes in complex networks. It involves predicting missing or fake links in static networks, or future and disappearing links in evolving networks\cite{liben2003link, shang2019link}, and forecasting links associated with newly introduced nodes\cite{ran2022predicting}. In social sciences, link prediction algorithms assist in identifying potential friendships\cite{benchettara2010supervised}. In biology, these algorithms are helpful for uncovering unknown protein-protein interactions\cite{oyetunde2017boostgapfill,amiri2022novel}. In computer engineering, link prediction helps in recommendation systems\cite{li2013recommendation} and knowledge graph completion \cite{nickel2015review}.

The problem of link prediction originated in the field of computer science \cite{liben2003link} and gained prominence within the domain of complex networks \cite{lu2011link}. Consequently, prior link prediction algorithms, which predominantly comprised heuristic approaches, were formulated in accordance with the intricate topological structures of complex networks and placed a greater emphasis on examining and interpreting their statistical physics characteristics \cite{lu2015toward,shang2019link,shang2022link}. 
These heuristic methods generally utilize properties such as closed triangular structures \cite{liben2003link,lu2010link}, clustering properties \cite{liu2016degree} and degree heterogeneity \cite{shang2019link} at the micro-level, communities \cite{yan2012finding,shang2020novel,ran2025machine} and long-line structure \cite{shang2022link} at the meso-level, and all paths between nodes \cite{katz1953new} and the average distance \cite{klein1993resistance,Fouss2007} at the global level. 


Heuristic algorithms identify network structural features well but struggle with the effective aggregation of node-specific information to enhance prediction accuracy. Subsequently developed network embedding algorithms \cite{Zhiyuli2016,Zhang2016,Zhangada2021} addressed this issue by synthesizing the local network structure surrounding nodes. While Graph Neural Networks (GNNs) \cite{wu2020comprehensive}, 
emerging later and rooted in a similar theoretical foundation, further refined this approach by introducing more sophisticated hierarchical information processing capabilities. These advances have proven effective for addressing the complexities of real-world networks, such as friendship networks, retweet networks, and academic paper citation networks, which typically contain a large number of nodes, each associated with various attributes like institution, age, gender, research field, and so on. Moreover, even though algorithms like ACT \cite{klein1993resistance,Fouss2007,lu2010link} and the Katz index \cite{katz1953new} attempt to rely on global structure for basic structural information aggregation, they suffer from high computational complexity with time complexities, significantly limiting their applicability to large-scale and attribute-rich networks. 

On the other hand, traditional machine learning approaches initially focused on a single level of structural information, such as local neighborhood patterns, global topological features, or quasi-global connectivity \cite{al2006link}. Subsequently, methods evolved to incorporate multiple topological similarities as features\cite{Chang2016,Yuan2021}, which is arguably the most intuitive approach. However, these methods still heavily rely on manually crafted heuristics. Consequently, they not only require significant human effort but also depend on domain-specific knowledge, thereby limiting their generalizability across different fields.

Fortunately, graph networks are capable of capturing topological information from the local to the global level, while simultaneously utilizing the inherent attributes of network nodes as initial features\cite{scarselli2008graph,battaglia2018relational,Cai2021}. Furthermore, Graph Convolutional Networks (GCNs) have gained recognition for their ability to integrate local nodes features with interpretable graph topology in convolutional layers, demonstrating significant potential in automatically capture higher-order information \cite{kipf2016semi}. This integration allows them to capture both the structural properties of the graph and the attributes of individual nodes, enabling a comprehensive understanding of complex relationships within networks. Importantly, their design supports scalable learning by efficiently aggregating information across neighborhoods, even in large-scale graphs. This capability makes them particularly effective for link prediction tasks in massive networks, where traditional methods often struggle due to computational complexity or limited feature utilization \cite{lei2019gcn,li2022temporal,shu2022link}.



Although GCNs utilize feature aggregation to incorporate both multi-level topological structures and intrinsic node attributes, the non-linear combination of multi-hop neighbor features during the message-passing process lacks interpretability. As a result, GCNs are often regarded as black-box models with limited physical or semantic meaning. To address this limitation, we extend the current GCNs framework by incorporating two fundamental physical properties of complex networks: triadic closure \cite{adamic2003friends,newman2001clustering} and heterogeneity \cite{barabasi1999emergence}. This enhancement provides a more foundational and general theoretical basis for GCNs, enabling higher algorithmic accuracy through previous rigorous analysis \cite{shang2019link,shang2022link,lu2011link},and validation across nine widely recognized datasets spanning complex networks and computer science domains.

\begin{figure*}[htbp]
\centering
\includegraphics[width=0.75\textwidth]{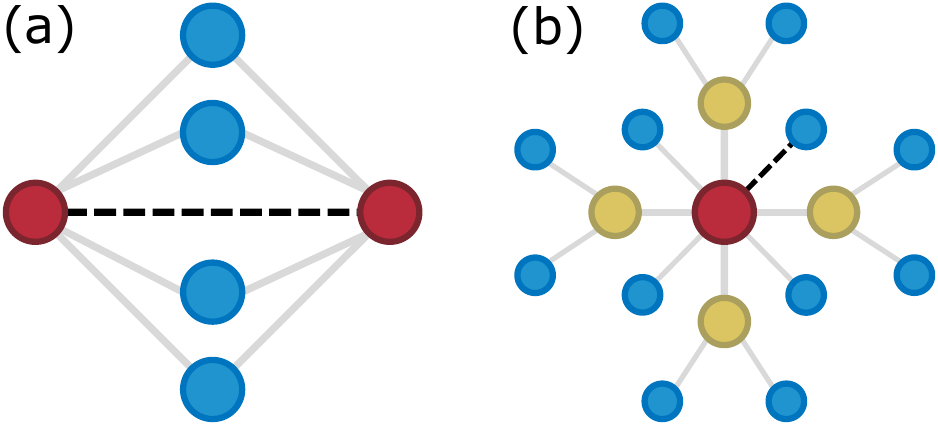}
\caption{In (a), it is observed that two red nodes share a significant number of blue nodes as their common neighbors. This indicates a high level of overlap in their immediate connections within the network. When a dashed line is introduced to directly link the two red nodes, multiple closed triangular structures are formed. This phenomenon aligns with the sociological principle of triadic closure, which posits that individuals in a social network tend to form connections with each other if they share mutual acquaintances or common neighbors. Triadic closure is a fundamental mechanism underlying the clustering behavior observed in many real-world networks. (b) highlights an issue related to structural heterogeneity within the network. Specifically, the connectivity pattern demonstrates a hierarchical decrease in the degree distribution as we move from the red nodes to the yellow nodes, and subsequently to the blue nodes. The degree of a node, defined as the number of edges connected to it, decreases progressively across these layers. Furthermore, there is a pronounced disparity between the node with the highest degree (often referred to as a hub node) and the nodes with minimal connectivity, such as leaf nodes. This structural heterogeneity reflects the uneven distribution of connections within the network, where hub nodes serve as central points of interaction, while peripheral nodes exhibit significantly lower levels of connectivity.}
\label{lpp}
\end{figure*}

Specifically, as depicted in Fig.~\ref{lpp}, we improve the feature aggregation mechanism of GCNs by introducing two explicit topological indicators --- common neighbor \cite{liben2003link} and degree difference between node pairs \cite{shang2019link} --- and propose a novel model named Triadic Closure-Heterogeneity-Harmony GCN (TriHetGCN). Moreover, a notable limitation of many traditional link prediction networks is the absence of node attributes, making them incompatible with standard GCNs that rely on such features for propagation and aggregation. To overcome this challenge, we compute shortest path distances to selected anchor nodes to extract structural features. This strategy not only improves the ability of the model to perceive node position and network structure but also enables direct processing of commonly used networks in link prediction tasks. Meanwhile, within the field of computer science, networks commonly used for link prediction inherently possess genuine node attributes. Our model can also directly utilize these intrinsic properties, enabling effective processing without the need for additional attribute extraction.

In other words, our method bridges the gap between traditional link prediction methods that focus on topological features and modern GNN-based methods that aggregate multi-level structural information through node attributes. TriHetGCN effectively integrates principles inspired by statistical physics with deep learning-based representations.
In many real-world networks, TriHetGCN achieves state-of-the-art performance and proposes a constructive framework for merging statistical physics and deep learning. Specifically, it demonstrates significant improvements in various applications, highlighting its versatility and effectiveness. The main contributions of this work are as follows:
\begin{enumerate}
    \item We introduce a new approach called TriHetGCN into the link prediction task, incorporating physical rules and nodes attributes into the framework to enhance the performance of traditional GNN-based models.
    \item For networks without node attributes, we propose a strategy to construct pseudo-node attributes using the shortest path distances to anchor nodes. This further strengthens the ability of TriHetGCN to capture both structural and positional information in the network.
    \item We explicitly incorporate degree difference and the number of common neighbors into the GCN aggregation mechanism. This design enhances the  ability of model to aggregate features from structurally significant node pairs, which are characterized by either a high number of shared neighbors or strong degree heterogeneity. In doing so, the model embeds two essential topological properties into the learning process and improves its interpretability.
    \item We conduct extensive experiments on nine widely used real-world datasets spanning complex networks to computer science. The results show that TriHetGCN achieves significantly higher link prediction accuracy than existing methods.
\end{enumerate}

The remainder of this paper is structured as follows. Section 2 provides a comprehensive review of the related work. In Section 3, we present a detailed description of our proposed model, TriHetGCN. Section 4 presents the experimental results and discusses their implications. Finally, Section 5 concludes the paper and outlines potential future research directions.

\section{Related work}
\label{sec:level2}

\subsection{Heuristics-based link prediction}
Early works on link prediction primarily focused on heuristic methods, which explored the topological information embedded in network graphs. These methods calculate node similarity scores to estimate the likelihood of links. Numerous such methods have been developed based on the triadic closure theory. Liben-Nowell and Kleinberg\cite{liben2003link} proposed several topological attributes such as common neighbors, Adamic-Adar and Katz Index, which have the ability to extract topological information to improve the performance of link prediction. 
Then, based on the indicators, some other topology-based predictors have been proposed for link prediction. For example, Zhou et al. utilized common neighbors, resource allocation (RA) and their degrees to infer the existence of potential links\cite{zhou2009predicting}. Lü et al. proposed a weighted link prediction algorithm by leveraging the actual edge weights and the theory of weak ties \cite{lu2010link}. Wu proposed a method that leverage local structure information by extracting the clustering coefficient of common neighbors \cite{wu2016link}. On the other hand, for topological information on heterogeneity. Shang\cite{shang2019link} discuss that heterogeneity is an obvious defining pattern for sufficiently sparse complex networks and incorporate heterogeneity information mechanism into the model online information propagation. Meanwhile, research on link prediction in temporal networks based on heuristic algorithms has gradually emerged \cite{lu2015toward,Shang2017}. Recent studies have also addressed the problem of predicting new edges introduced by newly added nodes \cite{ran2022predicting}, with the core principles still closely tied to triadic closure. 

Heuristic methods predict potential links through topological feature calculations, which are simple and have low time complexity. However, they rely heavily on the network topology, while ignoring node information or being unable to utilize large-scale, multi-dimensional node information, and only extract linear features. Therefore, they are insufficient for handling increasingly complex network prediction tasks.

\subsection{Learning-based link prediction}
In contrast to heuristic link prediction algorithms, learning-based algorithms, including machine learning and graph learning methods, that are capable of incorporating rich node attributes and capturing nonlinear relationships. Hasan et al. first considered link prediction as a binary classification task from a machine learning perspective\cite{al2006link}. By extracting similarity metrics between node pairs from network and treating them as features, the classifier can be constructed to distinguish between positive samples (forming links) and negative samples (not forming links). Since then, supervised classification methods have become prevalent in link prediction. For instance, Li et al. proposed a machine learning method that integrates logistic regression and XGBoost regularization model to obtain nonlinear information and features\cite{li2020ensemble}. Machine learning mentioned above are capable of extracting node features and capturing nonlinear information. However, for large-scale networks, the dimensions of the matrix are usually extremely large. The manual feature engineering for machine learning and network methods is labor-intensive and time-consuming. Moreover, hand-made features lack the ability to represent higher-order and more detailed information contained in the networks. To overcome those problems, we introduce graph deep learning methods into link prediction. 

Since graph deep learning methods have a strong ability in processing graph-structured data and obtaining higher-order features automatically, they are widely popular in various fields, such as recommender systems\cite{wang2021graph}, social network analysis\cite{wu2020comprehensive}.  As for link prediction, some existing research works have take advantage of graph neural netwrok and its variants, such as Graph Convolutional Network(GCN)\cite{kipf2016semi} and Graph Attention Network(GAT)\cite{velickovic2017graph}.
Hamilton et al.\cite{hamilton2017inductive} proposed a method called GraphSAGE employing diverse aggregator methods to extend GCN into an inductive learning framework, enabling the generation of node embeddings by aggregating neighborhood information.
Lei et al. took advantage of Graph Convolutional Network(GCN) models to achieve link prediction in complex networks by topological structure, and evolutionary structural features for predicting the missing link \cite{lei2019gcn}. 
Li et al. proposed a link prediction model based on GCN and the self-attention mechanism \cite{li2022temporal}. This model applied graph attention layers to capture the structural features of neighboring nodes and used graph convolutional layers to capture content features. 
Chen et al. integrated GCN into LSTM for link prediction to learn network structural and temporal information \cite{chen2022gc}. Their approach can utilize multimodal data to improve prediction accuracy in dynamic networks.
Shu et al. leveraged GCNs and the self-attention mechanism to separately extract the topological structural and temporal features of the network \cite{shu2022link}.

Graph deep learning methods can automatically learn multimodal features of the network, achieving better performance in link prediction compared to traditional methods. However, most existing graph deep learning approaches leverage GNNs and GCNs to obtain latent structural information. This approach can result in the loss of explicit network topology structural information based on physical theory analysis, which is clear and truly general, thereby affecting link prediction performance. Therefore, in this paper, we build upon our previous work on complex networks and link prediction to enhance the competitiveness and theoretical explainability of GCN methods.


\section{Methodology}
\label{sec:level3}

\subsection{Problem Statement}
We now introduce a formal description of our problem setting and preliminary knowledge to understand our work. The notations and symbols used are summarized in Table~\ref{Symbol Table}. 

\begin{table}[!ht]
    \centering
    \begin{tabular}{|c|c|}
    \hline
        Symbol & Description \\ \hline
        $U$ & A universal set which contains all possible node pairs \\ \hline
        $G$ & The real-world network data \\ \hline
        $v_i$ & The nodes of $G$ and $U$\\ \hline
        $e_{ij}$ & The link between $v_i$ and $v_j$ \\ \hline
        $E^P$ & Probe set \\ \hline
        $E^T$ & Training set \\ \hline
        $E^V$ & Validation set \\ \hline
        $A \in {\{0,1\}^{N\times N}}$ & The adjacency matrix of $G$ \\ \hline
        $d_i$ & The degree of the nodes $i$ \\ \hline
        $x_i$ & The feature of nodes $i$ \\ \hline
        $N$ & The number of network nodes \\ \hline
        $M$ & The number of anchor points \\ \hline
        ${dis}_j$ & The shortest path distance to the $j$-th anchor point \\ \hline
    \end{tabular}
    \caption{Symbol Table}
    \label{Symbol Table}
\end{table}

Let $G=(V,E,A,X)$ be an undirected static graph with $N$ nodes, where $V={v_1,v_2, \dots ,v_N}$ represents the set of nodes and $E=\{e_{ij}|v_i,v_j \in V\}$ is the set of observed edges where nodes $v_i,v_i \in V$ are connected.
The adjacency matrix $A\in R^{N \times N}$ is defined by $A_{ij}=1$ if there is an edge between node $v_i$ and node $v_j$, otherwise $A_{ij}=0$. 
The degree matrix $D={d_1,d_2, \dots ,d_N}$ of the graph is defined as the degree of each node. 
Besides, the node feature matrix is defined as $X = \{x_1, x_2, \dots, x_N\}$, where $x_i$ denotes the feature vector of node $i$. If the nodes in the network do not have inherent attributes, $x_i$ is constructed by computing the shortest path distances to a set of anchor nodes. These anchor nodes are selected as the top-ranked nodes with the highest degrees in the network. Otherwise, $x_i$ represents the inherent feature vector of node $i$.

No pair of nodes can have multiple links or self-links. It is assumed that some links within the static network are currently unobserved or missing. The goal of the link prediction task is to estimate the probability of a connection between two unlinked nodes by leveraging information inherent to the network.  

Typically, a certain proportion of edges from the network $G$ are selected to form the test set, while the remaining edges and all nodes constitute the training set. The training set is utilized to compute similarity scores between node pairs. In the field of complex networks, it is standard practice to allocate $10\%$ of the edges to the probe set $E^P$ and $90\%$ of the edges and nodes to the training set $E^T$. In contemporary computer science research,
the same proportion of edges is often designated for $E^P$.
However, $85\%$ of the edges and all nodes are typically assigned to $E^T$,
with the remaining $5\%$ reserved for a validation set $E^V$. This allocation allows for iterative model training, where $E^V$ serves to assess the performance of model after each training epoch. The model demonstrating the highest effectiveness on the $E^V$ is subsequently evaluated on $E^P$. To establish a consistent framework, we adopt the most widely-used dataset partitioning and testing standards in contemporary computer science.

In addition, we assume a universal set $U$, which contains all possible node pairs, regardless of whether a link exists between the nodes in each pair. For link prediction, the fundamental task is to ensure that the scores of existing links are higher than those of node pairs without links in $U\setminus E$. Hence in both testing, training, and validation processes, the positive samples consist of observed links, while the negative samples are randomly selected node pairs that do not have connections, with the same quantity as the number of existing links.

\subsection{Evaluation Metric}
We leverage two widely used evaluation metrics, Area Under Receiver Operating Characteristic Curve (AUC)  \cite{hanley1982meaning} and Average Precision (AP)\cite{aslam2006statistical,robertson2008new}, to assess the performance of our proposed model. Both AUC and AP are computed using functions provided by the standard Python library scikit-learn \cite{pedregosa2011scikit}.

\begin{enumerate}
\item AUC: Area Under Curve (AUC) evaluates the model’s ability to distinguish between positive and negative samples. It is defined as the area under the Receiver Operating Characteristic (ROC) curve, which plots the true positive rate (TPR) against the false positive rate (FPR) across different classification thresholds. Formally,
\begin{equation}
\mathrm{AUC}=\int_{0}^{1}TPR\left(x\right) d\left(\mathrm{FPR}\left(x\right)\right).
\end{equation}

Here, the true positive rate (TPR) and false positive rate (FPR) are defined as follows:

\begin{equation}
\mathrm{TPR}\ =\ \frac{\mathrm{TP}}{\mathrm{TP}\ +\ \mathrm{FN}},\ \mathrm{FPR}\ =\ \frac{\mathrm{FP} }{\mathrm{FP}\ +\ \mathrm{TN}},
\end{equation}
where TP, FP, FN, and TN are defined by the following:
TP: the number of correctly predicted positive samples.
FP: the number of negative samples incorrectly predicted as positive.
FN: the number of positive samples incorrectly predicted as negative.
TN: the number of correctly predicted negative samples.

\item AP: Average Precision (AP) summarizes the trade-off between precision and recall over all classification thresholds. It is defined as the area under the Precision-Recall (PR) curve. Formally,
\begin{equation}
\mathrm{AP}=\sum_{n=1}^{N}\left(R_n-R_{n-1}\right)\cdot P_n,
\end{equation}
where $P_n$ and $R_n$ represent the precision and recall at the n-th threshold, respectively, which are given by:
\begin{equation}
\mathrm{P}=\frac{\mathrm{TP}}{\mathrm{TP}+\mathrm{FP}},\ \mathrm{R}=\frac{\mathrm{TP}}{\mathrm{TP}+\mathrm{FN}}.
\end{equation}
\end{enumerate}

\subsection{Model overview}
The basic idea is to predict the possible link between two nodes by combining statistic physics and graph convolutional network, which can learn the topology structure representation of graph network for accurate link prediction. In this section, we present the construction of our proposed framework for link prediction based on GGNs and heuristic learning method.

\begin{figure*}[htbp]
\centering
\includegraphics[width=\textwidth]{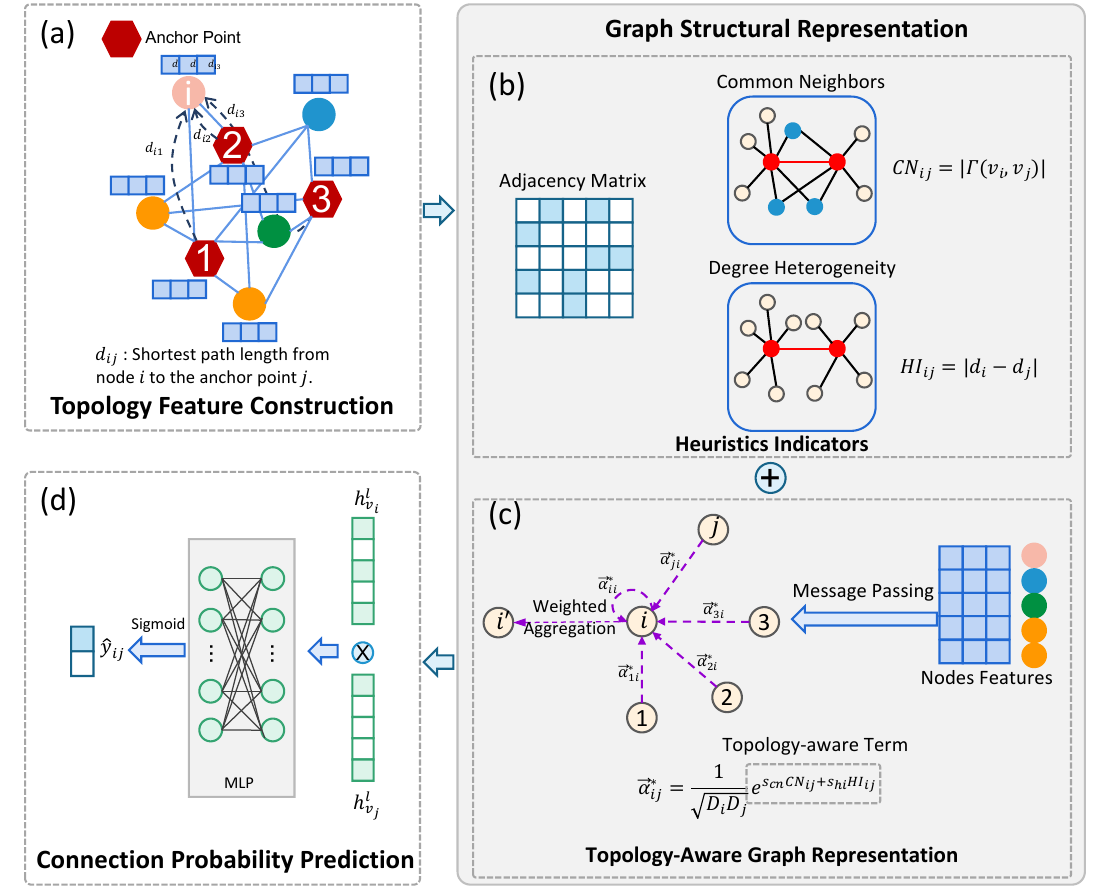}
\caption{An overview of the proposed TriHetGCN framework, which integrates physical statistic rules and GCN for link prediction, is presented as follows: (a) \textbf{Topology Feature Construction}: This module constructs node attributes based on the shortest paths from each node to $M$ selected anchor points, leveraging the underlying network structure. (b) \textbf{Heuristics Indicators}: This component utilizes physical rules, specifically the Common Neighbors (CN) and Heterogeneity Index (HI) within the network's topological structure, to capture explicit local relationships between nodes.
(c) \textbf{Topology-Aware Graph Representation}: This module enhances the GCN framework by incorporating local pairwise node relationships into the convolutional operator, refining its ability to capture fine-grained structural details through interactions between neighboring nodes. It then employs the node aggregation mechanism of GCN to extract global graph structural features. (d) \textbf{Connection Probability Prediction}: This component uses the complete spatial features as input variables and predicts the likelihood of link connections via a fully connected neural network.}
\label{overview}
\end{figure*}

As shown in Fig.~\ref{overview}, the link prediction framework is divided into three main components: topology feature construction, graph structural representation and connection probability prediction. First, we select the $M$ anchor nodes of the graph with the highest degrees, and calculate the shortest paths from each node to these anchor nodes, and use them as the features of the nodes. 
Second, we explicitly calculated common neighbor number and the degree heterogeneity index to capture the structural topological features of node pairs. 
Moreover, by explicitly integrating these two topological features into the GCN, we enhanced the GCNs representation of two essential features of complex networks, triadic closure and heterogeneity, during information aggregation. 
Third, in connection probability prediction module, we utilize a deep learning network to compute the probability of two nodes connection.

\subsection{Topology feature construction}
For traditional link prediction methods in the field of statistical physics, the network usually only contains the topological structure without node attributes. 
In our approach, we construct node attributes based on the network structure, as illustrated in Algorithm~\ref{Construction of Structural Node Features using Anchor Nodes}.
Given a real-world graph, we first identify all connected components and calculate the size of each. 
These components are then sorted in descending order by size. We define the important components as those whose cumulative size accounts for at least 80\% of the total number of nodes in the network.
Within each important component, we rank the nodes by their degrees and select the top 15\% as anchor points.
To ensure balance, the number of anchor nodes selected from any single component is capped at 150; if the number exceeds this threshold, only the top 150 highest-degree nodes are retained. 
In total, we obtain M anchor nodes. Subsequently, for each node in the network, we compute its shortest path distances to all M anchor nodes. These distances are normalized and used to form the node’s feature vector. 
The feature representation for node $i$ is defined as:

\begin{equation}
x_i=[{dis}_1,{dis}_2, \dots ,{dis}_j, \dots ,{dis}_M],
\end{equation}
where $\mathrm{dis}_j$ denotes the shortest path distance between node $i$ and the $j$-th anchor node.

\begin{algorithm}
\caption{Construction of Structural Node Features using Anchor Nodes}
\label{Construction of Structural Node Features using Anchor Nodes}
\begin{algorithmic}[1]
\State \textbf{Input:} Graph $G = (\mathcal{V}, \mathcal{E})$; anchor selection rate $r \in (0, 1)$; Minimum component size $s_{\text{min}}$; Max anchors per component $m$
\State \textbf{Output:} Node feature matrix $\mathbf{X} \in \mathbb{R}^{|\mathcal{V}| \times k}$

\Statex
\State Compute connected components $\{C_1, \dots, C_t\}$ of $G$

\State for i=1, \dots ,t do:
\Statex \quad If $|C_i| \ge \max(s_{\text{min}}, \text{size at 80 percentile})$:
\Statex \quad\quad Select top $\lfloor r \cdot |C_i| \rfloor$ highest-degree nodes as anchors $A_i$
\Statex \quad\quad Limit $|A_i| \le m$
\Statex \quad\quad Add $A_i$ to global anchor set $A$

\State Initialize feature matrix $\mathbf{X} \in \mathbb{R}^{|\mathcal{V}| \times k}$ with $-1$

\State For each anchor node $a_j \in A$:
\Statex \quad Compute shortest path distances $d(v, a_j)$ for all $v \in \mathcal{V}$
\Statex \quad Set $\mathbf{X}[v, j] \gets d(v, a_j)$

\State For each column $j$ in $\mathbf{X}$:
\Statex \quad Normalize: $\mathbf{X}[:, j] \gets \mathbf{X}[:, j] / \max(\mathbf{X}[:, j])$
\Statex \quad Replace $-1$ entries with $1 + \varepsilon$

\State \Return $\mathbf{X}$
\end{algorithmic}
\end{algorithm}

\subsection{Graph structural representation}
We built a combined heuristic learning algorithm and GCN framework to learn both local and global graph structural information.
The heuristic learning algorithm contains common neighbor and degree heterogeneity between node pairs, which capture the local structural information of the graph. 
As we all known, the GCN has a natural ability to capture global topology infomation of the network. 

First, we calculate the Common Neighbors index (CN)\cite{liben2003link}, which is a traditional local link prediction algorithm based on the closed triangular structure in the network and is used to measure the likelihood of a connection between nodes.
If two nodes have common neighbors, the likelihood of a link is relatively high. 
For any two nodes $v_i$ and $v_j$, the connection probability between nodes is measured by the number of their common neighbors, which is formulated as:

\begin{equation}
{CN}_{ij}=|\mathrm{\Gamma}(v_i,v_j)|,
\end{equation}
where $\mathrm{\Gamma}(v_i,v_j)$ represents the set common neighbors of nodes $v_i$ and $v_j$.

Second, the heterogeneity index (HI)\cite{shang2019link} is the difference quantification of the connection information in the network structure.
It reflects the different connection patterns between nodes and serves as a measure of network heterogeneity.
Since node degrees vary across the network, the heterogeneity index between node pairs reflects the non-uniformity of the connection structure, which is calculated as:
\begin{equation}
{HI}_{ij}=|d_i-d_j|,
\end{equation}
where $d_i$ and $d_j$ is the degree of nodes $v_i$ and $v_j$.

Finally, in this module, we consider two essential structural properties of complex networks—triadic closure and heterogeneity—during the GCN information aggregation process. To explicitly incorporate these structural cues, we integrate the common neighbor (CN) score and heterogeneity index (HI) of node pairs into the GCN aggregation mechanism. Furthermore, we introduce two learnable parameters, $s_{\mathrm{cn}}$ and $s_{\mathrm{hi}}$, which respectively control the contribution of CN and HI. These parameters are automatically optimized during training, allowing the model to adaptively adjust the structural influence for improved performance.

\begin{equation}
\phi_{ij}=\left(D_{ii}^{-1/2}D_{jj}^{-1/2}\right)\cdot\exp{\left(s_{\mathrm{cn}}\cdot{\mathrm{CN}}_{ij}+s_{\mathrm{hi}}\cdot{\mathrm{HI}}_{ij}\right)}.
\end{equation}

Here, $\phi_{ij}$ represents the normalized message-passing weight from node j to node i, combining standard GCN normalization with structural modulation. The overall graph convolution operation is then defined as:

\begin{equation}
H^{\left(1\right)}=\sigma\left(\Phi XW^{\left(0\right)}+b^{\left(0\right)}\right), H^{\left(l\right)}=\sigma\left(\Phi H^{\left(l-1\right)}W^{\left(l-1\right)}+b^{\left(l-1\right)}\right),
\end{equation}

where $X\in R^{N\times d}$ is the input node feature matrix, $H^l={h_{v_1}^l,h_{v_2}^l, \dots ,h_{v_N}^l}$ represents the current structure state of the graph, l is the number of GCN layers, $\Phi\in R^{N\times N}$ is the structure-aware propagation operator incorporating CN and HI, $W^{\left(l-1\right)}$ and $b^{\left(l-1\right)}$ are trainable parameters of the $l-1$th weights and bias, respectively. And $\sigma$ is the activation function. This combining method improve the weight of large heterogeneity index and common neighbor to enhance the performance of GCN.

\subsection{Connection probability prediction}
In order to calculate the similarity between two nodes, we select the feature vectors with structural information obtained in the previous step and compute their Hadamard product. The similarity between node pair $\left(v_i,v_j\right)$ is defined as:

\begin{equation}
{Similarity}_{ij}=h_{v_i}^l \otimes h_{v_j}^l
\end{equation}

Then, we apply two fully connected layer for computing the connection probability of two nodes, which can be defined as:

\begin{equation}
H=Mish\left(W_1\cdot\left({\rm Similarity}_{ij}\right)+b_1\right),
\end{equation}

\begin{equation}
\widehat{y_{ij}}=\sigma\left(W_2\cdot H+b_2\right),
\end{equation}
where $\mathrm{Mish}\left(\cdot\right)$ is the Mish activation function, and $\sigma\left(\cdot\right)$ denotes the sigmoid function. The output $\widehat{y_{ij}}\in\left(0,1\right)$ represents the predicted probability of a link between nodes $v_i$ and $v_j$.
$W_1\in R^{d\times h}$, $W_2 \in R^{h\times 1 }$ are the weight matrices of the decoder, and $b_1\in R^h$, $b_2\in R$ are the corresponding biases. Here, d is the embedding size, and h is the decoder hidden dimension.

Finally, we adopt the binary cross-entropy loss, which is widely used for binary classification tasks, to optimize the model. In each training epoch, the model predicts link existence probabilities for both existing edges and non-existing edges generated via negative sampling. The loss function is defined as:
\begin{equation}
Loss=-\sum_{\left(i,j\right)\in \varepsilon^+}\log{\left(\widehat{y_{ij}}\right)}-\sum_{\left(i,j\right)\in \varepsilon^-}\log{\left(1-\widehat{y_{ij}}\right)}\ ,
\end{equation}
where $\varepsilon^{+}$ and $\varepsilon^{-}$ are the sets of existing edges and non-existing edges respectively and $\widehat{y_{ij}}$ is the predicted probability of a link between nodes $i$ and $j$.

\section{Experiment}
\label{sec:level4}

\subsection{Datasets}
We conduct extensive experiments to verify our proposed model TriHetGCN, and evaluate the effectiveness of our model with night large-scale public datasets containing six real-world datasets provided by PyG Datasets and three synthetic graph datasets. 

The first series datasets are Cora\cite{sen2008collective}, Citeseer\cite{sen2008collective}, PubMed\cite{yang2016revisiting}, and DBLP\cite{bojchevski2017deep}- the citation datasets. In these datasets, each node represents a paper, and edges represent citation relationships between papers. Their node attributes are constructed by processing the textual content of each paper. Specifically, the preprocessing steps are as follows: First, all textual content from each paper is collected, followed by stemming and stop-word removal. Words appearing fewer than 10 times across the entire corpus are then removed, forming a final vocabulary list. For Cora, Citeseer, and DBLP, the node attributes are represented by a binary bag-of-words model, indicating whether each word in the vocabulary appears (1) or does not appear (0) in the corresponding paper. In contrast, the PubMed dataset uses the TF-IDF weighting scheme, resulting in continuous numeric attribute values.

The CS\cite{shchur2018pitfalls} dataset is a co-author dataset, where each node represents an author, and edges indicate co-authorship relations. Specifically, an edge exists between two nodes if the corresponding authors have co-authored at least one paper. The attributes of each author node are generated from the keywords of all papers authored by that individual. More precisely, all keywords from the papers authored by a given author are aggregated, and the node attributes are represented using a binary bag-of-words model.

The Facebook\cite{leskovec2012learning} dataset contains multiple independent sub-networks. For each sub-network, a reference point (known as ego) is selected for constructing the network. Nodes in the sub-network are the ego's friends. Edges form between nodes if they're friends with each other. The attributes of each Node in each ego-network are based on differences in profile information between node and ego, which represented by a binary value (0 for no difference, or 1 for difference compared to the ego).

We also select three commonly used networks for link prediction tasks based on statistical physics approaches: Power, Twitter, and INT. All three datasets contain only topological information and lack node features. The Power\cite{watts1998collective} dataset represents the topological structure of the power grid in the western United States. Nodes correspond to generators, transformers, and substations, and edges represent high-voltage transmission lines connecting these facilities. The Twitter\cite{de2013anatomy} dataset is a social network dataset, where each node represents a Twitter user, and edges represent retweet relationships between these users. The INT \cite{spring2004measuring} dataset maps router-level Internet topologies of real-world Internet Service Providers (ISPs). In this dataset, nodes represent routers, and edges indicate physical or logical connections between them. 
For networks like Power, Twitter and INT that lacks node features, we construct artificial node attributes for each node based solely on the network topology. The construction process is illustrated in FIG. 1. Specifically, we select a set of anchor nodes from large connected components based on node degrees. The anchor selection ratio  is set to 0.15, with a minimum component size threshold  corresponding to the 80th percentile of all component sizes. In addition, we limit the number of anchor nodes in each component to at most . For each node, the shortest path distance to each anchor node is computed. These distances are then normalized and used as node feature vectors.

The statistics of our dataset are shown in Table~\ref{Basic statistic of our datasets}. For all datasets, we choose the first 85\% of graphs for training data, 5\% for validation, and the 10\% for testing. And the negative samples are generated by the built-in RandomLinkSplit function provided in the PyTorch Geometric library.

\begin{table}[!ht]
    \label{Basic statistic of our datasets}
    \centering
    \begin{tabular}{ccccc}
    \hline
        Dataset & Number of nodes & Number of edges & Density & Nodes attributes \\ \hline
        Cora & 2708 & 5278 & 0.00144 & 1433 \\
        Citeseer & 3312 & 4660 & 0.00085 & 3703 \\
        PubMed & 19717 & 44327 & 0.00023 & 500 \\
        DBLP & 17716 & 52867 & 0.00034 & 1639 \\
        CS & 18333 & 81894 & 0.00049 & 6805 \\
        Facebook & 4039 & 88234 & 0.01082 & 1283 \\
        Power & 4941 & 6594 & 0.00054 & 150 \\
        Twitter & 256491 & 327374 & 0.00001 & 150 \\
        INT & 26848 & 41262 & 0.00011 & 600 \\ \hline
    \end{tabular}
    \caption{Basic statistic of our datasets}
\end{table}

\subsection{Experimental settings}

\textbf{Anchor nodes selection}: The network may not be fully connected and instead contains several subnetworks. 
For large subnetworks (with a cumulative size accounting for more than 80\% of the total network size), we select the top 15\% of the nodes with the highest degrees in these subnetworks as the anchor nodes of the network.

\textbf{Parameter settings}: For implementing these baselines, we adopt the parameter configurations recommended in their original implementations.
For GCN, we perform hyperparameter tuning using grid search to obtain the best performance.
The learning rate is selected from 0.001, 0.002, 0.003, 0.005, 0.01, and the number of hidden channels is chosen from 128, 256, 384. 
The best settings of these two parameters for each dataset are shown in Table~\ref{The parameter settings of learning rate and hidden channels for each dataset} after hyperparameter tuning. 
The number of training epochs is fixed at 1000, with an early stopping patience of 500. 
The dropout rate is set to 0.1. 
For each dataset, TriHetGCN uses the same model hyperparameters as GCN for the shared parameter settings. 
The initial values of the CN ratio and HI ratio are randomly selected from the range [0, 0.5], and their learning rate are independently set to 0.001.

\begin{table}[!ht]
    \centering
    \begin{tabular}{ccc}
    \hline
        Dataset & Learning rate & Hidden channels \\ \hline
        Cora & 0.01 & 128 \\
        Citeseer & 0.01 & 384 \\
        PubMed & 0.005 & 128 \\
        DBLP & 0.01 & 128 \\
        CS & 0.001 & 384 \\
        Facebook & 0.003 & 384 \\
        Power & 0.001 & 256 \\
        Twitter & 0.005 & 256 \\ 
        INT & 0.005 & 256 \\ \hline
    \end{tabular}
    \caption{The parameter settings of learning rate and hidden channels for each dataset}
    \label{The parameter settings of learning rate and hidden channels for each dataset}
\end{table}

\subsection{Performance Comparison}
We compare TriHetGCN with two categories of widely-used baseline models: (1) traditional heuristic-based link prediction methods, including Common Neighbors (CN)\cite{liben2003link}, Adamic-Adar (AA)\cite{adamic2003friends}, Resource Allocation (RA)\cite{zhou2009predicting}, Katz Index\cite{katz1953new}, and Local Path Index (LP)\cite{lu2009similarity} that are rooted in the triadic closure principle, as well as Random Walk with Restart (RWR)\cite{brin1998anatomy} and Local Random Walk (LRW)\cite{liu2010link} based on random walk theory; and (2) state-of-the-art learning-based graph neural network models, including Graph Convolution Networks (GCNs)\cite{kipf2016semi}, GraphSAGE\cite{hamilton2017inductive}, and Graph Attention Netwrok(GATs)\cite{velickovic2017graph}. 
As shown in Table~\ref{AUC results of link prediction on our datasets} (AUC) and Table~\ref{AP results of link prediction on our datasets} (AP), TriHetGCN consistently outperforms all baseline methods across all nine datasets, demonstrating stable and superior performance overall.

Compared to traditional heuristics, TriHetGCN achieves substantial improvements on multiple datasets. For instance, on Cora, Citeseer, and PubMed, local methods such as CN, AA, and RA yield AUC scores in the range of 64\%–72\%, while TriHetGCN reaches 93.69\%, 97.15\%, and 97.21\%, respectively. Among heuristics, Katz Index performs the best, with AUC scores of 83.76\%, 77.84\%, and 81.91\% on these datasets, but it still lags behind TriHetGCN by 10\%–20\%. Katz’s advantage mainly comes from its ability to model global distance through exponential path decay, which indirectly supports the effectiveness of our position-related embedding design. On networks without node attributes such as Power, Twitter, and INT, the performance gap becomes even more pronounced. Specifically, on Power, the AUC of Katz Index is 66.29\%, while TriHetGCN achieves 94.24\%; on Twitter, it improves from 54.04\% to 94.21\%; and on INT, from 72.67\% to 98.43\%. These results further demonstrate that replacing static path-count scores with shortest-path-based embeddings, which are integrated and optimized within a GCN framework, offers stronger expressive and generalization capabilities, especially in sparse and heterogeneous networks.

Compared to learning-based GNN models, TriHetGCN also maintains a consistent advantage across all datasets. On Cora, Citeseer, and PubMed, which contain rich node attributes, TriHetGCN achieves slightly higher AUC scores than GCN (e.g., 93.69\% vs. 92.73\% on Cora), indicating stable and reliable performance. Similar trends are observed on DBLP and CS. On featureless datasets, GraphSAGE and GAT perform poorly (e.g., 50.83\% and 52.44\% AUC on Power), while both GCN and TriHetGCN maintain strong performance. Notably, TriHetGCN consistently outperforms GCN, suggesting that the additional structural information introduced in our model—particularly triangle-based and heterogeneity-aware signals—further enhances representation capacity beyond the original GCN architecture.

\begin{table}[!ht]
    \centering
    \begin{tabular}{cccccccccc}
    \hline
        & Cora & Citeseer & PubMed & DBLP & CS & Facebook & Power & Twitter & INT \\ \hline
        CN & 71.94 & 67.04 & 64.39 & 77.48 & 89.86 & 99.16 & 58.01 & 53.32 & 62.75 \\
        AA & 71.44 & 66.86 & 64.44 & 77.48 & 89.89 & 99.26 & 58.05 & 53.31 & 62.84 \\
        RA & 71.90 & 67.13 & 64.42 & 77.57 & 89.88 & 99.39 & 58.06 & 53.32 & 62.77 \\
        Katz Index & 83.76 & 77.84 & 81.91 & 88.92 & 95.36 & 99.09 & 66.29 & 54.04 & 72.67 \\
        RWR & 82.63 & 75.58 & 75.24 & 87.46 & 95.02 & 98.50 & 62.90 & 47.76 & 68.82 \\
        LP & 80.58 & 75.27 & 79.92 & 86.85 & 94.05 & 99.12 & 62.65 & 60.82 & 72.67 \\
        LRW & 76.66 & 71.63 & 79.28 & 85.61 & 91.77 & 99.11 & 58.52 & 64.68 & 72.26 \\ \hline
        GCN & 92.73 & 96.55 & 97.18 & 96.30 & 97.84 & 99.50 & 93.72 & 93.93 & 98.33 \\
        GraphSAGE & 92.21 & 95.17 & 94.32 & 95.62 & 96.51 & 99.44 & 50.83 & 91.58 & 95.67 \\
        GAT & 92.14 & 96.39 & 95.64 & 95.96 & 97.32 & 99.28 & 52.44 & 75.37 & 93.11 \\
        TriHetGCN & \textbf{93.69} & \textbf{97.15} & \textbf{97.21} & \textbf{96.34} & \textbf{97.92} & \textbf{99.52} & \textbf{94.24} & \textbf{94.21} & \textbf{98.43} \\ \hline
    \end{tabular}
    \caption{AUC results of link prediction on our datasets. Due to the large scale of the Twitter dataset, the Katz Index, Random Walk with Restart (RWR), Local Path Index (LP), and Local Random Walk (LRW) methods could not be directly applied to the full graph. To address this issue, we performed network sampling by randomly selecting a node with a fixed seed (42), and extracting all nodes within its 4-hop neighborhood. The resulting subgraph, containing 3,303 nodes and 4,711 edges, was used for evaluating these four traditional link prediction methods. In this paper, the heuristic algorithm is independently repeated $100$ times, while the GCNs-related algorithms are independently repeated $10$ times. Since we used the most popular random seed, $42$, in the field of computer science, when splitting the dataset into training, validation, and prob sets, our results are reproducible. A fixed random seed ensures consistent partitioning across different runs because pseudo-random number generators (PRNGs) produce sequences of numbers based on an initial value—the random seed. When the same seed is set, the PRNG generates an identical sequence of numbers, which in turn determines how the samples are allocated to the training, validation, or prob sets.}
    \label{AUC results of link prediction on our datasets}
\end{table}

\begin{table}[!ht]
    \centering
    \begin{tabular}{cccccccccc}
    \hline
        & Cora & Citeseer & PubMed & DBLP & CS & Facebook & Power & Twitter & INT \\ \hline
        CN & 71.72 & 66.96 & 64.36 & 77.43 & 89.81 & 98.96 & 57.98 & 53.24 & 62.75 \\
        AA & 71.53 & 66.88 & 64.45 & 77.51 & 89.91 & 99.16 & 58.02 & 53.45 & 62.87 \\
        RA & 72.00 & 67.15 & 64.43 & 77.59 & 89.89 & 99.36 & 58.02 & 53.46 & 62.79 \\
        Katz Index & 84.32 & 78.08 & 82.71 & 89.59 & 95.84 & 98.96 & 66.28 & 57.29 & 73.10 \\
        RWR & 88.01 & 80.58 & 84.80 & 91.69 & 96.65 & 97.98 & 77.82 & 63.11 & 73.35 \\
        LP & 80.56 & 75.25 & 79.93 & 86.95 & 94.14 & 98.98 & 62.62 & 57.90 & 72.70 \\
        LRW & 77.07 & 71.75 & 79.40 & 85.79 & 91.93 & 98.80 & 58.45 & 65.64 & 72.34 \\ \hline
        GCN & 93.51 & 97.02 & 97.18 & \textbf{96.71} & 98.07 & 99.44 & 92.27 & 95.24 & 98.06 \\
        GraphSAGE & 93.09 & 96.13 & 95.29 & 96.28 & 97.04 & 99.43 & 55.15 & 93.13 & 94.55 \\
        GAT & 93.13 & 96.59 & 95.07 & 95.91 & 97.51 & 99.06 & 51.37 & 70.20 & 91.39 \\
        TriHetGCN & \textbf{94.40} & \textbf{97.53} & \textbf{97.24} & \textbf{96.71} & \textbf{98.19} & \textbf{99.46} & \textbf{92.58} & \textbf{95.45} & \textbf{98.17} \\ \hline
    \end{tabular}
    \caption{AP results of link prediction on our datasets}
    \label{AP results of link prediction on our datasets}
\end{table}

\subsection{Ablation Study}
We conduct ablation studies to evaluate the performance improvements brought by TriHetGCN.
To evaluate the contribution of each module in TriHetGCN, we implement the following variants:

\textbf{GCN+CN}: 
To demonstrate the effectiveness of the common neighbor, we incorporate the number of common neighbors between node pairs into the message-passing mechanism of GCN.

\textbf{GCN+HI}: 
To demonstrate the effectiveness of the degree difference, we incorporate the degree difference between node pairs into the message-passing mechanism of GCN.

The experimental results are presented in Table~\ref{AUC results of ablation study} (AUC) and Table~\ref{AP results of ablation study} (AP). 
We observe that both GCN+CN and GCN+HI outperform the vanilla GCN baseline across most datasets, indicating that each structural feature provides improvements in link prediction. 
Notably, TriHetGCN, which integrates both CN and HI with learnable weights, consistently achieves the best performance, confirming the complementary benefits of combining these two heuristics. 

To better understand why CN or HI performs better on specific datasets, we further examine the structural characteristics of each network, focusing on two dimensions: triangle structure and degree heterogeneity. The triangle structure is quantified by the global clustering coefficient, which is calculated as the number of closed triplets multiplied by three, divided by the number of open triplets\cite{newman2001clustering}, while degree heterogeneity is measured using the coefficient of variation of node degrees.

\begin{figure*}[htbp]
\centering
\includegraphics[width=0.9\textwidth]{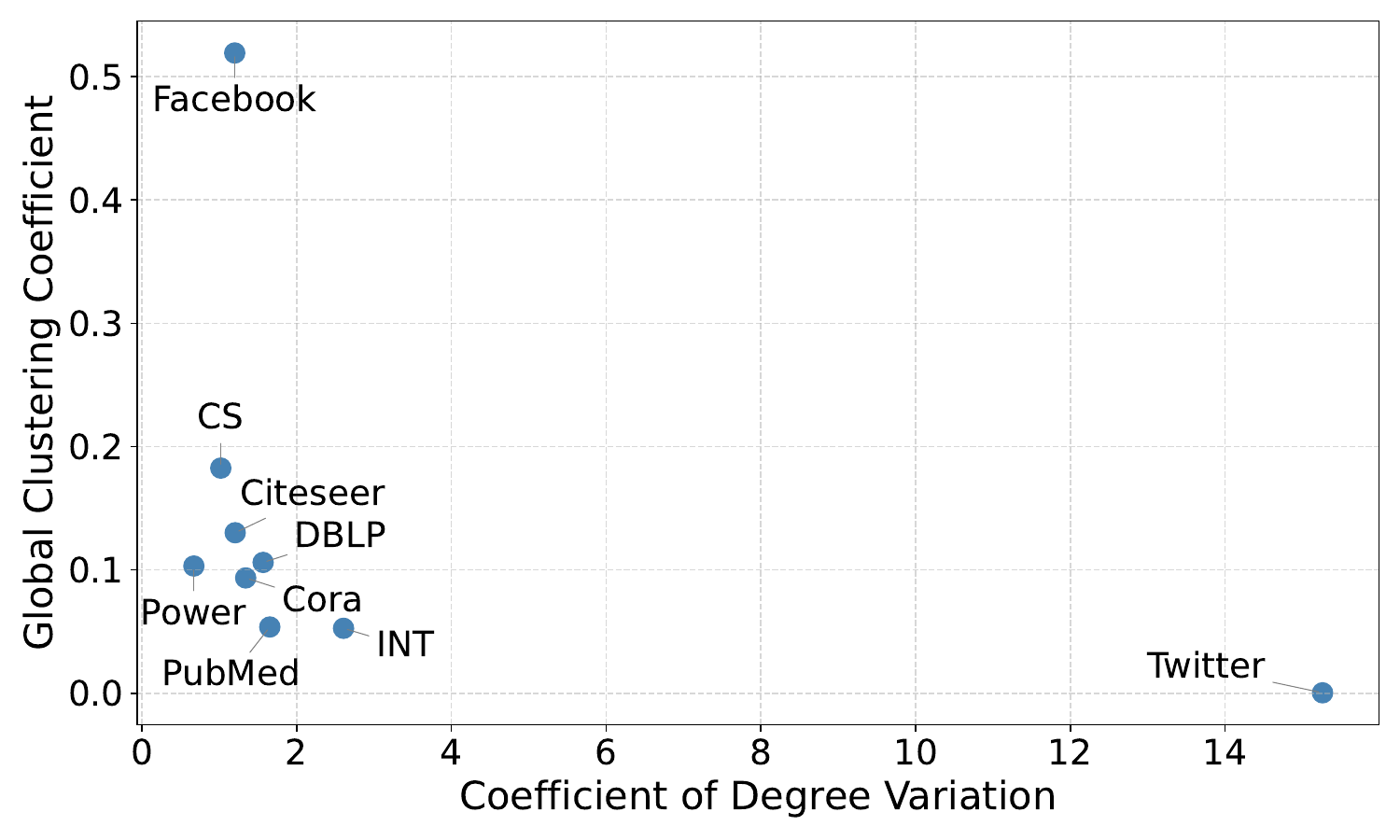}
\caption{Network Heterogeneity and Triangle Structure: The distribution of these two structural properties across selected datasets is illustrated. Degree heterogeneity is measured by the coefficient of degree variation (x-axis). The common neighbor property is measured by the average clustering coefficient, which is proportional to the number of closed triangles in the network (y-axis). Each point represents a dataset, labeled by its name.}
\label{Network Heterogeneity and Triangle Structure}
\end{figure*}

As shown in Fig.~\ref{Network Heterogeneity and Triangle Structure}, different datasets exhibit distinct structural features. For example, Facebook and CS networks have relatively high clustering coefficients but low degree heterogeneity, suggesting a strong presence of triangle closures. In these networks, the CN-based variant (GCN+CN) achieves improvements, as the common neighbor heuristic aligns well with the local clustering patterns. In contrast, Twitter shows extremely high degree heterogeneity and a very low clustering coefficient, indicating a hub-dominated and loosely clustered structure. In such cases, the HI-based variant (GCN+HI) performs better, as it captures the influence of connectivity disparity among node pairs. For datasets like Cora, Citeseer, and DBLP, which exhibit moderate values of both clustering and heterogeneity, both CN and HI contribute meaningfully, and the full model (TriHetGCN) shows notable performance advantages.

\begin{table}[!ht]
    \centering
    \begin{tabular}{cccccccccc}
    \hline
        & Cora & Citeseer & PubMed & DBLP & CS & Facebook & Power & Twitter & INT \\ \hline
        GCN & 92.73 & 96.55 & 97.18 & 96.30 & 97.84 & 99.49 & 93.72 & 93.93 & 98.33 \\
        GCN+CN & 93.04 & 96.74 & 97.19 & 96.33 & 97.86 & 99.50 & 94.08 & 94.17 & 98.36 \\
        GCN+HI & 93.47 & 96.78 & 97.20 & \textbf{96.39} & 97.86 & 99.51 & 94.13 & 94.16 & 98.39 \\
        TriHetGCN & \textbf{93.69} & \textbf{97.15} & \textbf{97.21} & 96.34 & \textbf{97.92} & \textbf{99.52} & \textbf{94.24} & \textbf{94.21} & \textbf{98.43} \\ \hline
    \end{tabular}
    \caption{AUC results of ablation study}
    \label{AUC results of ablation study}
\end{table}

\begin{table}[!ht]
    \centering
    \begin{tabular}{cccccccccc}
    \hline
        & Cora & Citeseer & PubMed & DBLP & CS & Facebook & Power & Twitter & INT \\ \hline
        GCN & 93.51 & 97.02 & 97.18 & 96.71 & 98.07 & 99.42 & 92.27 & 95.24 & 98.06 \\
        GCN+CN & 93.71 & 97.23 & 97.01 & \textbf{96.72} & 98.03 & \textbf{99.48} & \textbf{92.64} & 95.38 & 98.11 \\
        GCN+HI & 94.08 & 97.16 & \textbf{97.25} & \textbf{96.72} & 98.10 & 99.44 & 92.60 & 95.41 & 98.11 \\
        TriHetGCN & \textbf{94.40} & \textbf{97.53} & 97.24 & 96.71 & \textbf{98.19} & 99.44 & 92.58 & \textbf{95.45} & \textbf{98.17} \\ \hline
    \end{tabular}
    \caption{AP results of ablation study}
    \label{AP results of ablation study}
\end{table}

\section{Conclusion}
\label{sec:level5}
In this paper, we introduce a novel model named TriHetGCN, designed to address the limitations of traditional heuristic algorithms and existing GNN-based approaches. By explicitly incorporating topological indicators, such as the number of common neighbors and degree differences, into the node information propagation and aggregation process, TriHetGCN enhances representation learning for each node. Additionally, we propose a method to construct pseudo-node attributes based on shortest path distances to selected anchor nodes, enabling our model to function effectively even in social networks lacking inherent node attributes. This advancement extends the applicability of TriHetGCN across various real-world datasets, demonstrating superior performance in link prediction tasks both with and without intrinsic node attributes.

Despite its promising results, there are two primary areas for future improvement. First, calculating shortest path distances from all nodes to selected anchor nodes presents a significant computational challenge, with a time complexity of $O(MN)$, where $M$ is the number of anchor nodes and $N$ is the total number of network nodes. Future work will focus on strategies to optimize computational efficiency while retaining global structural awareness. Second, while our current framework is tailored for static networks, extending TriHetGCN to dynamic graphs---where network structures evolve over time with new nodes, as in our previous work \cite{ran2022predicting}---presents an exciting avenue for further research.

\section*{Data Available}
The cleaned usable data from the public dataset and the reproducible code for this article are available at \url{https://github.com/wordbomb/TriHetGCN} (to be made public upon acceptance of the article).

\bibliographystyle{unsrt}  
\bibliography{main}

\end{document}